\documentclass[aps,prl,twocolumn,showpacs,superscriptaddress,bibliography]{revtex4-1}
\usepackage{graphicx}    
\usepackage{dcolumn}    
\usepackage{bm}             
\usepackage{amssymb}   
\usepackage{amsmath}    
\usepackage{subfigure}    
\usepackage{braket}
\usepackage{mathrsfs}
\usepackage{multirow}
\usepackage[export]{adjustbox}
\usepackage[toc,page]{appendix}
\begin{document}

\title{Individual addressing of trapped ion qubits with geometric phase gates}

\author{R. T. Sutherland}
\email{robert.sutherland@utsa.edu}
\affiliation{Department of Electrical and Computer Engineering, University of Texas at San Antonio, San Antonio, Texas 78249, USA}
\author{R. Srinivas}
\affiliation{Department  of  Physics,  University  of  Oxford,  Clarendon  Laboratory, Parks  Road,  Oxford,  OX1  3PU,  UK}
\author{D. T. C. Allcock}
\affiliation{Oxford Ionics Limited, Begbroke Science Park, Begbroke, OX5 1PF, UK}
\affiliation{Department of Physics, University of Oregon, Eugene, Oregon 97403, USA}

\date{\today}

\begin{abstract}
We propose a new scheme for individual addressing of trapped ion qubits, selecting them via their motional frequency. We show that geometric phase gates can perform single-qubit rotations using the coherent interference of spin-independent and (global) spin-dependent forces. The spin-independent forces, which can be generated via localised electric fields, increase the gate speed while reducing its sensitivity to motional decoherence, which we show analytically and numerically. While the scheme applies to most trapped ion experimental setups, we numerically simulate a specific laser-free implementation, showing cross-talk errors below $10^{-6}$ for reasonable parameters.
\end{abstract}
\maketitle

To achieve universal error-corrected quantum computation, we require one- and two-qubit gates with infidelities below $\sim 10^{-5}$ and $\sim 10^{-4}$, respectively \cite{nielsen_2010,lidar_2013}. Perhaps more challenging, we must also perform them in a system that can scale to thousands of qubits. Trapped ions are one of the most promising platforms for achieving these requirements \cite{cirac_1995, monroe_1995, wineland_1998, haffner_2008,blatt_2008}, due to their high gate fidelities, long coherence times, and all-to-all connectivity \cite{ladd_2010,harty_2014,ballance_2016,gaebler_2016,srinivas_2021,clark_2021}. Even so, there are challenges to address before the platform can reliably operate with enough qubits to perform useful computations. The now demonstrated quantum charge-coupled device (QCCD) \cite{home_2009, pino_2020}, where electrodes are used to move ions between separated trap `zones', each with designated functions, could provide the modularization needed to meet this challenge \cite{wineland_1998,kielpinski_2002}. The question remains, however, as to the best way of generating the gate fields themselves.

\begin{figure}[b]
\includegraphics[width=0.5\textwidth]{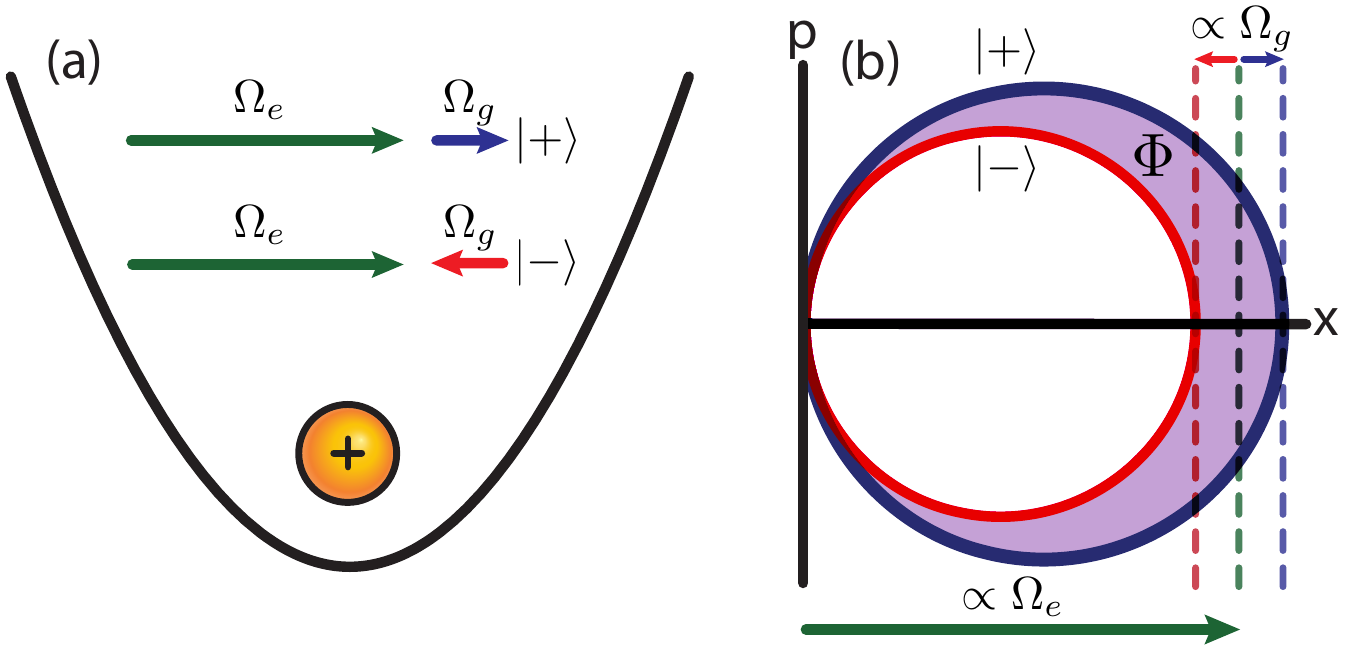}
\centering
\caption{Illustration of the proposed single-qubit geometric phase gate. (a) An E-field, with Rabi frequency $\Omega_{E}$, is combined with a similarly detuned spin-dependent force, with Rabi frequency $\Omega_{p}$, producing constructive and destructive interference with the $\ket{+}$ (blue) and $\ket{-}$ (red) eigenstates of the gate's Pauli operator $\hat{\sigma}_{\alpha}$, respectively. (b) This interference causes the eigenstates to have distinct phase-space trajectories. The difference in areas (purple) of the closed trajectories ${\Phi \equiv \Phi_{+}-\Phi_{-}}$ corresponds to the gate's rotation angle.}
\label{fig:fig_1}
\end{figure}

There are two general approaches to creating these fields, either with lasers or magnets. Laser-based approaches have achieved one-qubit and two-qubit gates with infidelities of $\sim 10^{-5}$, and $\sim 10^{-3}$ \cite{ballance_2016,gaebler_2016,clark_2021}, respectively, and are straightforwardly focused for individual addressing~\cite{naegerl_1999}. Unfortunately, they suffer from photon scattering~\cite{ozeri_2007}, as well as phase and amplitude noise. Laser-free gates, by contrast, do not suffer from these issues \cite{wineland_1998,mintert_2001,ospelkaus_2008,johanning_2009,ospelkaus_2011,timoney_2011,harty_2014,harty_2016,weidt_2016,lekitsch_2017,srinivas_2018, sutherland_2019,zarantonello_2019,sutherland_2020,srinivas_2021}. Similarly, they can operate at infidelities of $\sim 10^{-6}$ for one-qubit gates \cite{harty_2014}, and $\sim 10^{-3}$ for two-qubit gates \cite{harty_2016, zarantonello_2019, srinivas_2021}. Laser-free approaches typically use microwaves and magnetic field gradients, which cannot be localized to individual ions; this makes single-qubit addressing more difficult. 

The QCCD architecture makes this somewhat easier, as it enables gate implementations in zones that are separated by 100s of microns. However, at these separations there can still be significant cross-talk because magnetic fields from current-carrying electrodes only decrease polynomially with distance, and there can potentially be inductive coupling to other electrodes. While cross-talk can be mitigated somewhat with active cancellation fields \cite{aude_2014,aude_2017}, schemes that require fields resonant with qubit transitions will likely become more complicated with qubit number $N$. This has been avoided by separating the ions in \textit{qubit} frequency space using magnetic field gradients \cite{wang_2009, johanning_2009, warring_2013_prl, weidt_2016, srinivas_2021}. Similarly, one could create differential qubit frequency shifts between different zones in a QCCD architecture, but this is limited in two ways. First, the shifts should be stable/repeatable, requiring significant book-keeping for larger systems. Secondly, the Rabi frequencies of the gates must be smaller than the separations of their qubit frequencies which could create a speed limit for larger $N$. 

In this work, we propose the first geometric phase gate \cite{molmer_1999,molmer_2000,milburn_2000,leibfried_2003} operation for directly implementing single-qubit rotations\textemdash decreasing hardware complexity by enabling single qubit gates to be implemented with the same fields used for two-qubit gates. Like its two-qubit parallel, our scheme enables the separation of ions in \textit{motional} frequency space by differentiating zone potentials with locally tuned electrodes. Thus, the scheme can be implemented exclusively with fields that are detuned from all qubit transitions without modifying their frequencies.

Our scheme operates via the interference of spin-dependent and spin-independent forces, both similarly detuned from the motion. Here, both eigenstates of the spin operator return to the origin in phase-space, but accumulate different geometric phases (see Fig. \ref{fig:fig_1}). The scheme has an effective Rabi frequency $\propto \sqrt{\Omega_{E}\Omega_{p}}$, where $\Omega_{E}$ is the Rabi frequency of an applied E-field, and $\Omega_{p}$ is that of an applied B-field gradient. The dependence on $\sqrt{\Omega_{\text{E}}}$ enables much faster single-qubit gates compared to equivalent two-qubit gates; it is much easier to generate large E-field couplings relative to those from magnetic field gradients. Compared to a $\propto \Omega_{p}$ two-qubit gate, we show that infidelities are decreased by at least a factor of $\sim \Omega_{p}/\Omega_{E}$ for static motional frequency shifts and motional dephasing, as well as $\sim (\Omega_{p}/\Omega_{E})^{3/2}$ for heating. Lastly, we perform numerical simulations using experimental parameters similar to Ref.~\cite{srinivas_2021}, showing we are able to reduce cross talk to below $10^{-6}$.

We want to generate a single-qubit rotation:
\begin{eqnarray}\label{eq:model_gate}
\hat{U}_{g} = e^{-i\theta \hat{\sigma}_{\alpha}/2},
\end{eqnarray}
where $\theta$ is the rotation angle of the gate while $\hat{\sigma}_{\alpha} \equiv \hat{n}\cdot \vec{\sigma}$, such that $\hat{n}$ is a unit vector, and $\vec{\sigma}=(\hat{\sigma}_{x},\hat{\sigma}_{y},\hat{\sigma}_{z})$ is a vector comprising the Pauli matrices. We will generate this effective interaction with accumulated geometric phases, where the eigenstates $\ket{\pm}$ of $\hat{\sigma}_{\alpha}$ (eigenvalues $\pm 1$) acquire distinct phases $\Phi_{\pm}$ (see Fig.~\ref{fig:fig_1}(b)). Representing the system in the $\ket{\pm}$ eigenbasis, this gives a time-propagator:
\begin{eqnarray}
\hat{U}_{p}\ket{\psi} &=& c_{+}e^{i\Phi_{+}}\ket{+} + c_{-}e^{i\Phi_{-}}\ket{-} \nonumber \\
&=& e^{i(\Phi_{+}+\Phi_{-})/2}\Big(c_{+}e^{-i\Phi/2}\ket{+} + c_{-}e^{i\Phi/2}\ket{-} \Big),~~~~~~
\end{eqnarray}
where $\Phi \equiv \Phi_{-}-\Phi_{+}$, showing, up to a global phase, that $\hat{U}_{g}$ is equivalent to $\hat{U}_{p}$ when $\theta = \Phi$.

To generate $\hat{U}_{p}$, we add an E-field to the typical geometric phase gate interaction:
\begin{eqnarray}\label{eq:orig_gp}
\hat{H}_{p} &=& \hbar\Omega_{p}\hat{\sigma}_{\alpha}\hat{a}^{\dagger}e^{i\Delta t} + \hbar\Omega_{E}\hat{a}^{\dagger}e^{i[\Delta t+\phi]} + c.c.,
\end{eqnarray}
where $\Delta$ is the gate detuning, $\hat{a}^{\dagger}(\hat{a})$ is a creation(annihilation) operator, and $\phi$ is the phase of the E-field relative to the spin-motion coupling. We have considered Eq.~(\ref{eq:orig_gp}) in a frame rotating with respect to the qubit frequency, mode frequency $\omega_{r}$, and have assumed the rotating wave approximation for terms oscillating near these frequencies. It is here worth noting that Refs.~\cite{turchette_1998, leibfried_1999, warring_2013, warring_2013_prl} consider spin-dependent interactions coupled to the traps' rf micromotion, similarly operating via spin-motion and E-field terms; this scheme is disadvantageous compared to Eq.~(\ref{eq:orig_gp}), because it leads to temperature dependent, and non-zero higher-order, terms in the Magnus expansion \cite{supplemental}. Furthermore, using the micromotion requires pushing ions off the rf null to induce transitions. Aside from being sensitive to stray E-fields which can move the ions and lead to unwanted cross talk, this requires an extra transport operation, which can be slow.
We can exactly describe $\hat{U}_{p}$ with the Magnus expansion \cite{magnus_1954} up to $2^{\text{nd}}$-order:
\begin{eqnarray}\label{eq:magnus}
\hat{U}_{p} \!= \exp\!\Big(\!-\frac{i}{\hbar}\int^{t_{g}}_{0}\!\!\!\!dt^{\prime}\hat{H}_{p}(t^{\prime}) -\frac{1}{2\hbar^{2}}\!\!\int^{t_{g}}_{0}\!\!\!\!\int^{t^{\prime}}_{0}\!\!\!\!\!dt^{\prime}dt^{\prime\prime}\Big[\hat{H}_{p}(t^{\prime}),\hat{H}_{p}(t^{\prime\prime})\Big]\Big), \nonumber \\
\end{eqnarray}
where higher-order terms vanish. We ensure the phase space trajectories of $\ket{\pm}$ close by setting $t_{g}=2\pi K/\Delta$, where $K$ is the number of loops traversed in phase-space. Up to a global phase this gives:
\begin{eqnarray}
\hat{U}_{p} &=& \exp\Big(\frac{4\pi i K\Omega_{p}\Omega_{E}}{\Delta^{2}}\cos[\phi]\hat{\sigma}_{\alpha} \Big),
\end{eqnarray}
showing the gate is most efficient when the two interactions in $\hat{H}_{p}$ are aligned, i.e. $\phi\in\{0,\pi\}$; we henceforth assume $\phi = 0$. Setting $\Delta = (8\pi K\Omega_{p}\Omega_{E}/\theta)^{1/2}$, we see $\hat{U}_{p}=\hat{U}_{g}$, up to a global phase, with a gate time $t_{g} = (\pi K\theta/2\Omega_{p}\Omega_{E})^{1/2}$, and an effective Rabi frequency of $\Omega_{\text{eff}} \equiv \theta/t_{g} = (2\Omega_{p}\Omega_{E}\theta/\pi K)^{1/2}$. Because its speed scales $\propto \sqrt{\Omega_{E}\Omega_{p}}$, it can potentially operate orders of magnitude faster than than a $\propto \Omega_{p}$ two-qubit gate. All of the interactions used to generate the gate are tuned to the frequency of the addressed well, affecting a gate with potentially no fields on-resonant with any spectator qubit transitions; this enables the elimination of cross-talk with pulse-shaping. 

While this derivation considered a single ion in a well, it can be extended straightforwardly to include multiple ions in a well. This could, in theory, be used to address individual ions in the same well \cite{supplemental}. This requires the ability to address a complete set of $N$ orthogonal modes of motion, which could make the technique challenging to implement when many ions are in the same well. In contrast to B-field gradients, addressing some modes with an E-field requires a differential component; for example, the coupling strength to the stretch mode of a two-ion crystal is proportional to the difference in the E-field amplitudes at the position of each ion.

\begin{figure}[b]
\includegraphics[width=0.5\textwidth]{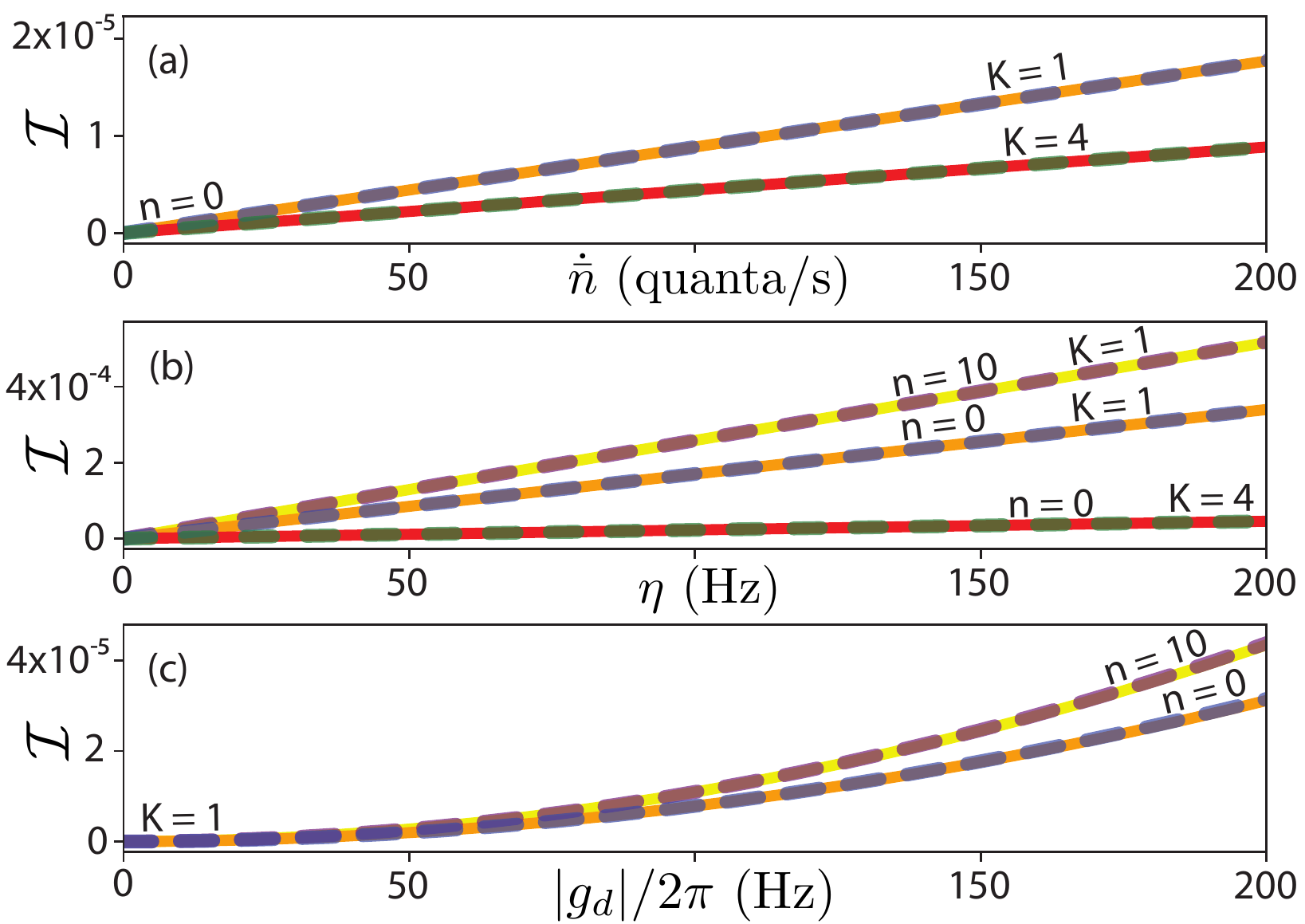}
\centering
\caption{Infidelity $\mathcal{I}$ versus (a) heating rate $\dot{\bar{n}}$ (b) dephasing rate $\eta$ and (c) static motional frequency shift magnitude $|g_{d}|$ for analytic (dashed) and numerical (solid) calculations, when $\Omega_{p}/2\pi=2~\text{kHz}$ and $\Omega_{E}/2\pi = 100~\text{kHz}$. In parts (a-c), we show calculations initialized to the motional ground state $n=0$ with $K=1$ loops in phase space. In parts (a) and (b), we illustrate that increasing $K$ decreases the effect $\dot{\bar{n}}$ or $\eta$ has on $\mathcal{I}$, showing calculations where $K=4$ and $n=0$. In parts (b) and (c), we illustrate the scheme's temperature dependence versus $\eta$ and $g_{d}$, showing calculations where $K=1$ and $n=10$. Note that when $\mathcal{I}\ll 1$, $\mathcal{I}$ is symmetric about $g_{d}=0$.}
\label{fig:motional}
\end{figure}

Trapped ion quantum computers traditionally operate in a regime where the infidelity is significantly lower for one-qubit gates than for two-qubit gates. This is because two-qubit gates are slower than one qubit gates, and they require a temporal window where the qubits are entangled to the motion, rendering them sensitive to motional decoherence as well as added control errors \cite{molmer_1999,molmer_2000,sutherland_2022_1}. As we are now proposing a scheme for one-qubit gates that \textit{also} requires spin-motion entanglement, it is important to ensure that motional decoherence affects $\mathcal{I}$ significantly less than its two-qubit counterpart. We here consider the contribution of heating, motional dephasing, and static motional frequency shifts to the gate infidelity $\mathcal{I}$. In order to evaluate $\mathcal{I}$, we follow a technique that dates back to NMR \cite{haeberlen_1968}, but has been used more recently to evaluate the contribution of various sources of noise to $\mathcal{I}$ for two qubit gates \cite{haddadfarshi_2016,martinez_2021,sutherland_2022_1}. We consider a Hamiltonian in the presence of an error term:
\begin{eqnarray}
\hat{H}_{t} = \hat{H}_{p} + \hat{H}_{e},
\end{eqnarray}
where $\hat{H}_{p}$ is given by Eq.~(\ref{eq:orig_gp}). Here, $\hat{H}_{e}$ represents the error term to be considered where $\hat{H}_{e}=2\hbar g_{h}\cos(\omega t)(\hat{a}^{\dagger}e^{i\omega_{r}t} + \hat{a}e^{-i\omega_{r}t})$ for heating and $\hat{H}_{e}=\hbar g_{d}\cos(\omega t)\hat{a}^{\dagger}\hat{a}$ for dephasing, where $g_{h(d)}$ are Rabi frequencies. For each source of error, we evaluate the infidelity by first transforming into the interaction picture with respect to either $\hat{H}_{e}$ (heating) or $\hat{H}_{p}$ (motional dephasing and motional frequency shifts), in order to produce a factorized time propagator $\hat{U}_{t,\omega} = \hat{U}_{p}\hat{U}_{e,\omega}$, making the final equation for infidelity:
\begin{eqnarray}\label{eq:infidelity}
\mathcal{I}_{\omega} &=& \sum_{n^{\prime}}|\bra{\psi(0)}\bra{n^{\prime}}\hat{U}_{e,\omega}\ket{\psi(0)}\ket{n}|^{2}, 
\end{eqnarray}
We evaluate $\hat{U}_{e,\omega}$ either by Taylor expanding the exponential (heating) or using perturbation theory (motional dephasing), both up to $2^{\text{nd}}$-order. Finally, we determine $\mathcal{I}$ by averaging over $S_{\omega}$, the normalized spectral power density:
\begin{eqnarray}
\mathcal{I} = \int^{\infty}_{0}d\omega S_{\omega}\mathcal{I}_{\omega},
\end{eqnarray}
subsequently making the Born-Markov approximation \cite{sutherland_2022_1,supplemental}. As shown in Ref.~\cite{sutherland_2022_1}, this is mathematically equivalent to the Lindblad formalism when $\mathcal{I}\ll 1$. We also determine $\mathcal{I}$ for static motional frequency shifts, i.e. the limit $S_{\omega}\rightarrow \delta(\omega)$ where $\delta(\omega)$ is the Dirac delta function, negating the need for the Born-Markov approximation \cite{sutherland_2022_1,supplemental}.

\begin{table}[t]
\begin{center}
\begin{tabular}{| c | c |}
\hline
\textbf{Error type} & \textbf{Effect on gate fidelity} \\
\hline\hline
 heating & $\mathcal{I} = \dot{\bar{n}}\sqrt{\frac{\Omega_{p}\theta^{3}}{8 \pi K \Omega_{E}^{3}}}\lambda^{2}_{\hat{\sigma}_{\alpha}}$\\ 
 \hline
 motional dephasing & $\mathcal{I} = \eta\sqrt{\frac{\Omega_{p}\theta^{3}}{32\pi K\Omega_{E}^{3}}}\Big(2n+1 + \frac{3\theta\Omega_{E}}{2\pi K\Omega_{p}}\Big)\lambda^{2}_{\hat{\sigma}_{\alpha}}$\\  
 \hline
 motional frequency & \multirow{2}{*}{$\mathcal{I}  = \frac{g_{d}^{2}\theta^{2}}{16\Omega_{E}^{2}}\Big(2n+1 + \frac{2\theta\Omega_{E}}{\pi K \Omega_{p}} \Big)\lambda_{\hat{\sigma}_{\alpha}}^{2}$} \\ shifts & \\
 \hline 
\end{tabular}
\caption{Table of infidelities $\mathcal{I}$ for systems with motional states with an initial Fock state $n$, due to common sources of motional decoherence.}\label{table:errors}
\end{center}
\end{table}

The results of these calculations\textemdash recorded in Table~\ref{table:errors}\textemdash show that we are proposing one qubit gates that are significantly less sensitive to motional decoherence than their two-qubit counterparts. Compared to the analytic formulae for two-qubit gates in Ref.~\cite{sutherland_2022_1}, we find that $\mathcal{I}$ is reduced by at least a factor of $(\Omega_{p}/\Omega_{E})^{3/2}$ for heating, and of $\Omega_{p}/\Omega_{E}$ for motional dephasing and static motional frequency shifts. Since $\Omega_{E}$ is due simply to an E-field interacting with a charge, it can be several orders-of-magnitude larger than $\Omega_{p}$ (potentially limited by the validity of the rotating wave approximation), leading to an increase in speed and insensitivity to motional errors. This insensitivity to motional decoherence is further illustrated in Fig.~\ref{fig:motional}, where we show $\mathcal{I}$ versus heating rate $\dot{\bar{n}}$, motional decoherence rate $\eta = 2/\tau = \pi\varepsilon^{2} S_{0}/2$, where $\tau$ is the coherence time between neighboring Fock states, and static motional frequency shift magnitude $|g_{d}|$. Here, we compare the equations presented in Table~\ref{table:errors} to direct numerical integration; this figure clearly demonstrates the accuracy of our analytic formulae in the relevant parameter regimes. The decrease in sensitivity to these sources of motional decoherence is larger than expected from the $\propto \sqrt{\Omega_{E}/\Omega_{p}}$ increase to the gate speed. This is due to the fact that the increased speed comes from the interference of the gradient with the external E-field, increasing the gate speed while simultaneously decreasing the time-averaged spin-motion entanglement. 

\begin{figure}[b]
\includegraphics[width=0.5\textwidth]{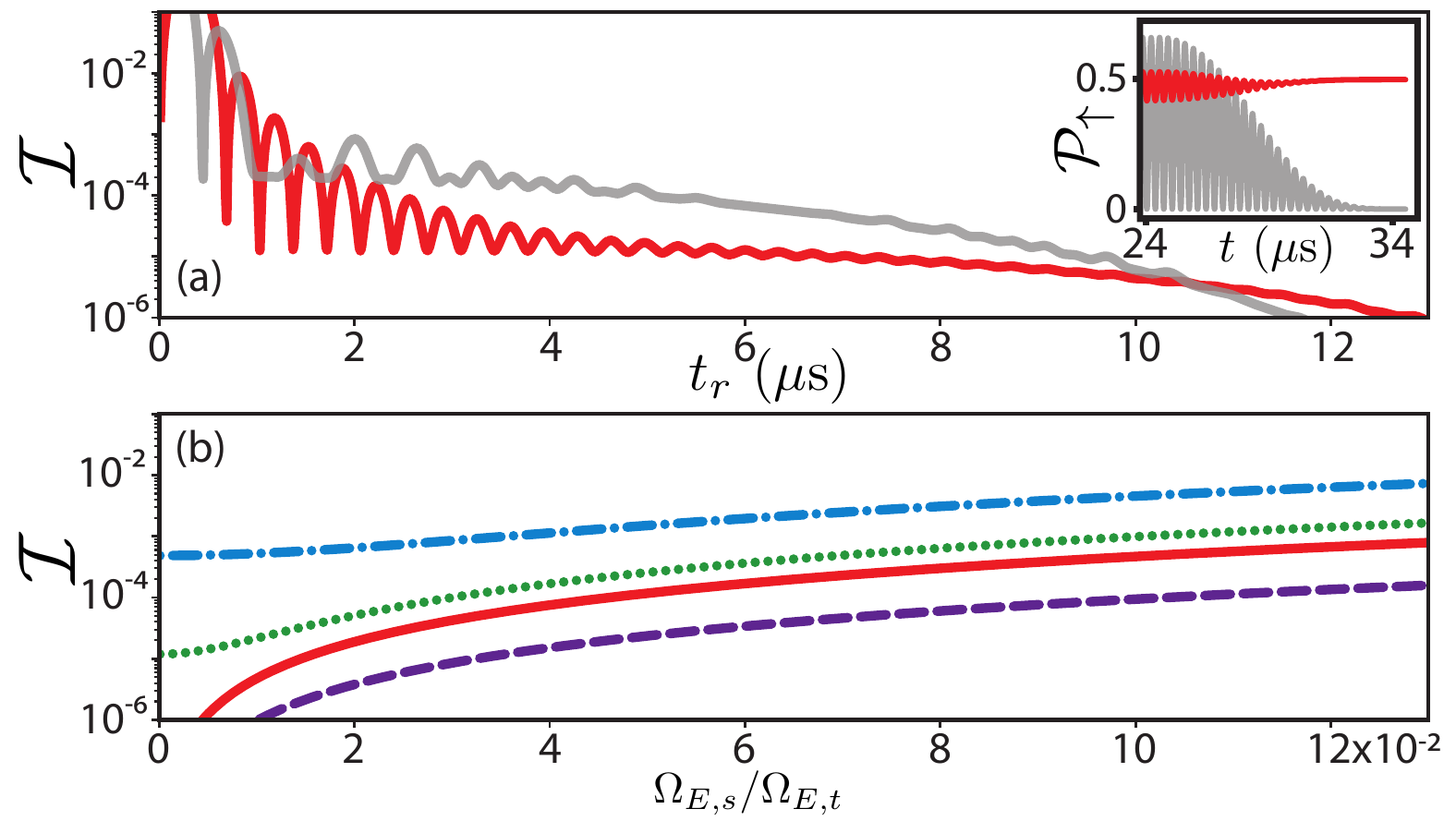}
\centering
\caption{ (a) Numerically calculated infidelity $\mathcal{I}$ of target qubit (red) and spectator qubit (grey) versus ramping time $t_{r}$ for an effective $\theta = \pi/2$ gate. Here, the the motional frequency of the target qubit is set to $\omega_{r,t}/2\pi = 6.5~\text{MHz}$, while the motional frequency of the spectator qubit is set to $\omega_{r,s}=6.3~\text{MHz}$. The inset shows the probability of measuring $\ket{\uparrow}$ for both qubits $\mathcal{P}_{\uparrow}=|\braket{\uparrow|\psi_{t,s}(t)}|^{2}$ when $t_{r}=10~\mu\text{s}$. (b) Infidelity of spectator qubit versus E-field amplitude $\Omega_{E,s}$ relative to the E-field seen by the target qubit $\Omega_{E,\tau}$ when $t_{r}=15~\mu\text{s}$. This is shown for the same motional frequency of the target qubit, and the motional frequency of the spectator qubit is set to $\omega_{r,s}/2\pi = 6.1~\text{MHz}$ (purple dashed), $\omega_{r,s}/2\pi = 6.3~\text{MHz}$ (red solid), $\omega_{r,s}/2\pi = 6.35~\text{MHz}$ (green dotted), and $\omega_{r,s}/2\pi = 6.4~\text{MHz}$ (blue dot dashed).}
\label{fig:cross_talk}
\end{figure}

Finally, we evaluate our scheme's resilience to cross-talk with a specific physical implementation \cite{sutherland_2019,srinivas_2018,srinivas_2021}, noting that any protocol capable of generating a two-qubit geometric phase gate\textemdash and the necessary E-field\textemdash is capable of implementing it. We here consider two ions in separate wells, where one ion acts as the `target' qubit, and the other as the `spectator' qubit. The target well is designated a motional frequency $\omega_{\tau}$, while the spectator mode is designated a \textit{detuned} motional frequency $\omega_{s}$. Both are driven with a pair of bichromatic microwave fields with Rabi frequency $\Omega_{\mu}$, symmetrically detuned around the qubit frequency, and an rf B-field gradient with frequency $\omega_{g}/2\pi = 5~\text{MHz}$. The resulting interaction is 

\begin{eqnarray}\label{eq:microwave_ham}
\hat{H}_{\mu} &=& 2\hbar\sum_{\gamma=\tau,s}\Omega_{\mu}\hat{\sigma}_{+,\gamma}\cos(\delta t) + \Omega_{g}\hat{\sigma}_{z,\gamma}\cos(\omega_{g}t)\hat{a}^{\dagger}_{\gamma}e^{i\omega_{\gamma}t}\!\! \nonumber \\
&& + \Omega_{E,\gamma}\sin([\omega_{\tau}-\Delta]t)\hat{a}^{\dagger}_{\gamma}e^{i\omega_{\gamma}t} + c.c.,
\end{eqnarray}
where $\hat{\sigma}_{\alpha,\gamma}$ is a Pauli operator acting on qubit $\gamma$, $\Omega_{E,\gamma}$ is the E-field Rabi frequency seen by qubit $\gamma$, and $\delta$ is the detuning of the microwave fields from the qubit frequency. The above equation is in the rotating frame with respect to the motional and qubit frequencies, and we have made the rotating wave approximation for terms oscillating near the qubit frequency. Similar to Ref.~\cite{sutherland_2020}, we can set $\delta = (\omega_{\tau}-\omega_{g})-\Delta$, and the time-propagator resulting from Eq.~(\ref{eq:microwave_ham}) can be evaluated in the interaction picture with respect to the $\propto \Omega_{\mu}$ term. Dropping fast rotating terms, this gives:
\begin{eqnarray}\label{eq:interation_ham}
\hat{H}_{\mu,I} \simeq i\hbar\Big(\Omega_{g}J_{1}\Big[\frac{4\Omega_{\mu}}{\delta}\Big]\hat{\sigma}_{y,\tau} + \Omega_{E,\tau}\Big)\hat{a}_{\tau}^{\dagger}e^{i\Delta t} + \text{c.c.},~~
\end{eqnarray}
taking the form of Eq.~(\ref{eq:orig_gp}) acting on the target qubit. 

We now determine the validity of Eq.~(\ref{eq:interation_ham}), showing that with pulse shaping it becomes a very good approximation. As was shown in Ref.~\cite{sutherland_2019}, if we perform this gate while smoothly ramping $\Omega_{\mu(g)}$ on and off before and after the gate, respectively, then the time-propagator given by Eq.~(\ref{eq:interation_ham}) converges to the time-propagator for Eq.~(\ref{eq:microwave_ham}). This enables high-fidelity gates \cite{srinivas_2021}, despite the Rabi flopping due to the $\propto \Omega_{\mu}$ term. There is no pulse shaping for the E-fields in our simulations. In Fig.~\ref{fig:cross_talk}, we show this convergence for a $\theta = \pi/2$ gate, where $\omega_{\tau}/2\pi = 6.5~\text{MHz}$, $\Omega_{g}/2\pi = 2~\text{kHz}$, $\Omega_{E,\tau}/2\pi = 100~\text{kHz}$. The $\propto \Omega_{\mu(g)}$ fields are simultaneously ramped on and off according to $\sin^{2}(t/t_{r})$, ideally leading to $\hat{U}_{g}\rightarrow e^{i\pi\hat{\sigma}_{y,\tau}/4}$ for the target qubit and $\hat{U}_{g}\rightarrow \hat{I}$ for the spectator qubit. Figure \ref{fig:cross_talk}(a) shows that for values over $t_{r}\simeq 13~\mu\text{s}$, $\Omega_{E,s}=0$, and $\omega_{s}/2\pi = 6.3~\text{MHz}$, $\mathcal{I}$ can be suppressed to below $10^{-6}$ for both the target and spectator qubits. We have here optimized $t_{g}$ to maximize the fidelity of the target qubit operation when $t_{r}=15~\mu\text{s}$, which makes $t_{g}\simeq 40.2~\mu\text{s}$ when including the $2t_{r}$ of pulse shaping time. We found no limit to the degree to which the cross-talk can be reduced via increasing $t_{r}$ in our simulations, indicating a trade-off between reducing cross-talk and $t_{g}$. The inset of Fig.~\ref{fig:cross_talk} illustrates the time dynamics of the target and spectator qubits, showing that while both qubits experience high-frequency oscillations during the gate, after the pulse shaping sequence, both qubits converge to the final states predicted by Eq.~(\ref{eq:interation_ham}). Finally, while $\Omega_{E,s}$ will likely decrease rapidly with spatial separation, it will not vanish entirely. Because of this, Fig.~\ref{fig:cross_talk}(b) shows $\mathcal{I}$ versus $\Omega_{s,E}$; we here plot values of $\omega_{s}/2\pi = 6.1-6.4~\text{MHz}$, showing that when $\omega_{s}$ is further detuned from $\omega_{\tau}$, one should expect less cross-talk (keeping all other parameters fixed) due to the fact that all spectator transitions are more off-resonant.

In conclusion, we have proposed a new one-qubit geometric phase gate scheme that is much faster and more robust to noise than its two-qubit counterpart, while also enabling the suppression of cross-talk to an arbitrary degree\textemdash even with global qubit control fields. We first developed the theory for this gate sequence, showing it requires a standard set of interactions present in most trapped ion laboratories. We then showed, analytically and numerically, that our proposed one-qubit geometric phase gates are significantly less sensitive to motional decoherence compared to their two-qubit gate counterparts. Finally, we provided a numerical simulation of one (of many) possible physical implementations of our scheme, showing that when it is combined with pulse shaping it can reduce cross-talk to an arbitrary degree, even when both qubits experience the same microwave fields and gradient. This work is important to the prospects of scalable, laser-free trapped ion architectures because it shows how to perform targeted operations without spatially localized qubit control fields.

The authors would like to thank C. J. Ballance, and D. H. Slichter for helpful discussions. 

\bibliography{biblio}
\clearpage
\onecolumngrid
\appendix

\section{Addressing with rf micromotion}

As in Refs.~\cite{leibfried_1999, warring_2013_prl}, we consider a trapped ion system with a sideband interaction and an E-field force, potentially from the rf micromotion:
\begin{eqnarray}
\hat{H}_{\text{lab}} &=& \hbar\omega_{r}\hat{a}^{\dagger}\hat{a} + \frac{\hbar\omega_{0}}{2}\hat{\sigma}_{z} + 2\hbar\Omega_{\text{rsb}}\cos([\omega_{0}-\omega_{r}+\Delta]t)\hat{\sigma}_{x}\Big(\hat{a}^{\dagger} + \hat{a} \Big) + 2\hbar\Omega_{e}\cos([\omega_{r}-\Delta]t)\Big(\hat{a}^{\dagger} + \hat{a} \Big),
\end{eqnarray}
where $\omega_{0}$ is the qubit frequency, and $\omega_{r}$ is motional mode frequency. Transforming into the interaction picture with respect to the motional and qubit frequencies, as well as dropping counter-rotating terms, gives:
\begin{eqnarray}\label{eq:rsb_rot_ham}
\hat{H} &=& \hbar\Omega_{\text{rsb}}\Big(\hat{\sigma}_{-}\hat{a}^{\dagger}e^{i\Delta t} + \hat{\sigma}_{+}\hat{a}e^{-i\Delta t} \Big) + \hbar\Omega_{e}\Big(\hat{a}^{\dagger}e^{i\Delta t} + ae^{-i\Delta t} \Big).
\end{eqnarray}
We now plug Eq.~(\ref{eq:rsb_rot_ham}) into the Magnus expansion \cite{magnus_1954} while setting the gate time $t_{g} = 2\pi N/\Delta$, observing that the $1^{\text{st}}$-order term in the Magnus expansion sinusoidally integrates to zero. Dropping a global phase, this gives:
\begin{eqnarray}\label{eq:time_prop_rsb}
\hat{U} &=& \exp\Big(\frac{i t_{g}}{\Delta}\Big[\Omega_{\text{rsb}}^{2}\hat{\sigma}_{z}\Big\{\hat{a}^{\dagger}\hat{a} + \frac{1}{2} \Big\} + \Omega_{\text{rsb}}\Omega_{e}\hat{\sigma}_{x} \Big] \Big),
\end{eqnarray}
up to $2^{\text{nd}}$-order. Because $[\hat{H}(t^{\prime}),[\hat{H}(t^{\prime\prime}),\hat{H}(t^{\prime\prime\prime})]]\neq 0$, the expansion cannot be exactly expressed in a straightforward manner, which could limit the fidelity\textemdash even in ideal conditions. Equation~(\ref{eq:time_prop_rsb}) also reveals a temperature dependent $\propto \hat{a}^{\dagger}\hat{a}$ term, further limiting the approach's potential for high-fidelity operations.

\section{Single-qubit addressing for ions in the same well}

Assuming a rotating frame with respect to the motional and qubit frequencies, we consider a Hamiltonian describing the $j^{\text{th}}$ mode of a chain of $N$ ions driven with a geometric phase gate interaction and a similarly detuned E-field:
\begin{eqnarray}\label{eq:collective_geo}
\hat{H} &=& \hbar\Omega_{p,j}\hat{S}_{\alpha,j}\Big(\hat{a}^{\dagger}_{j}e^{i\Delta t} + \hat{a}_{j}e^{-i\Delta t} \Big) + \hbar\Omega_{e,j}\Big(\hat{a}_{j}^{\dagger}e^{i\Delta t} + \hat{a}_{j}e^{-i\Delta t} \Big),
\end{eqnarray}
where $\Omega_{p,j}$ and $\Omega_{e,j}$ are Rabi frequencies, and $\hat{S}_{\alpha,j}\equiv \sum_{n}e_{n,j}\hat{\sigma}_{\alpha,n}$ is a collective Pauli operator. Importantly, the contribution $e_{n,j}$ of the $n^{\text{th}}$ ion to $\hat{S}_{\alpha,j}$ is proportional to its projection onto mode $j$. Plugging Eq.~(\ref{eq:collective_geo}) into the Magnus expansion \cite{magnus_1954}, and evaluating at a time $t_{g}=2\pi K/\Delta $, gives a time propagator: 
\begin{eqnarray}
\hat{U}_{p}^{\prime} = \exp\Big(\frac{i t_{g}}{\Delta}\Big[\Omega_{p,j}^{2}\hat{S}_{\alpha,j}^{2} + 2\Omega_{p,j}\Omega_{e,j}\hat{S}_{\alpha,j}\Big] \Big). 
\end{eqnarray}
We can see that this operation leaves an extraneous $\propto \hat{S}^{2}_{\alpha,j}$ term, creating unwanted entanglement between the qubits; because $\hat{\sigma}_{\alpha}^{2}=\hat{I}$, this term corresponds to a global phase for one qubit systems. This entanglement term can be eliminated by dividing the operation into two $K/2$ loop operations while flipping the sign of the detuning $\Delta\rightarrow -\Delta$, and the sign of $\hat{S}_{\alpha,j}\rightarrow -\hat{S}_{\alpha,j}$ with a spin-echo sequence. Doing so results in a total time propagator:
\begin{eqnarray}\label{eq:collective_eff}
\hat{U}_{p} &=& \exp\Big(\frac{-i t_{g}}{2\Delta}\Big[\Omega_{p,j}^{2}\hat{S}_{\alpha,j}^{2} - 2\Omega_{p,j}\Omega_{e,j}\hat{S}_{\alpha,j}\Big] \Big)\exp\Big(\frac{i t_{g}}{2\Delta}\Big[\Omega_{p,j}^{2}\hat{S}_{\alpha,j}^{2} + 2\Omega_{p,j}\Omega_{e,j}\hat{S}_{\alpha,j}\Big] \Big) \nonumber \\
&=& \exp\Big(\frac{2 i\Omega_{p,j}\Omega_{e,j} }{\Delta}\hat{S}_{\alpha,j}t_{g} \Big) \nonumber \\
&=& \exp\Big(-i\frac{\theta}{2}\hat{S}_{\alpha,j}\Big),
\end{eqnarray}
giving a single-qubit gate with an effective rotation angle of $\theta\equiv -2\Omega_{p,j}\Omega_{e,j}/\Delta$.

Because $e_{n,j}$ is proportional to each ion's projection onto each mode, the ability to perform this operation on a complete set of $N$ modes should, in theory, give experimentalists the ability to perform individual qubit addressing. In a well with two ions, for example, if we can address a center-of-mass mode $\hat{S}_{\alpha,c} = \hat{\sigma}_{\alpha,2}+\hat{\sigma}_{\alpha,1}$ and a stretch mode $\hat{S}_{\alpha,s}=\hat{\sigma}_{\alpha,2}-\hat{\sigma}_{\alpha,1}$, we can address qubit $2$ by generating Eq.~(\ref{eq:collective_eff}) with an angle $\theta$ for each rotation:
\begin{eqnarray}
\hat{U}_{2} &=& \exp\Big(-i\frac{\theta}{2}\hat{S}_{\alpha,c} \Big)\exp\Big(-i\frac{\theta}{2}\hat{S}_{\alpha,s} \Big) \nonumber \\
&=& \exp\Big(-i\theta\hat{\sigma}_{\alpha,2} \Big),
\end{eqnarray}
corresponding to a $2\theta$ rotation on qubit 2. Similarly, performing two operations with opposite signed values of $\theta$ gives:
\begin{eqnarray}
\hat{U}_{2} &=& \exp\Big(i\frac{\theta}{2}\hat{S}_{\alpha,s} \Big)\exp\Big(-i\frac{\theta}{2}\hat{S}_{\alpha,c} \Big) \nonumber \\
&=& \exp\Big(-i\theta\hat{\sigma}_{\alpha,1} \Big),
\end{eqnarray}
giving a $2\theta$ rotation on qubit $1$. This process can, theoretically, be extended to chains of $N$ ions. In practice, however, it may be difficult to produce an E-field differential pronounced enough to generate large values of $\Omega_{e,j}$ in wells with many ions. We therefore focus the main body of this work on systems with one ion in a well, likely the most relevant to scalable QCCD architectures. 

\section{Infidelities from motional decoherence}

In this section, we discuss the effect of three sources of motional decoherence on the fidelity of the single-qubit geometric phase gates we proposed in the main body. First, in a frame rotating with respect to the qubit and trap frequencies, we examine the geometric phase gate Hamiltonian acting in the presence of an error Hamiltonian:
\begin{eqnarray}\label{eq:error_gen}
\hat{H}_{t} &=& \hat{H}_{p} + \hat{H}_{e}, \nonumber \\
&=& \hbar\hat{A}\Big(\hat{a}^{\dagger}e^{i\Delta t} + \hat{a}e^{-i\Delta t} \Big) + \hat{H}_{e},
\end{eqnarray}
where we have introduced $\hat{A}=\Omega_{p}\hat{\sigma}_{\alpha}+\Omega_{e}\hat{I}$, and have assumed we can apply the rotating wave approximation for $\hat{H}_{e}$. 

\subsection*{Static motional frequency shifts}

We first consider static motional frequency shifts $\hat{H}_{e}\equiv \hbar g_{d} \hat{a}^{\dagger}\hat{a}$, which gives:
\begin{eqnarray}
\hat{H}_{t} &=& \hbar\hat{A}\Big(\hat{a}^{\dagger}e^{i\Delta t} + \hat{a}e^{-i\Delta t} \Big) + \hbar g_{d}\hat{a}^{\dagger}\hat{a}.
\end{eqnarray}
For static motional frequency shifts and motional dephasing, we analyze the gate fidelity by following a technique laid out in Ref.~\cite{sutherland_2022_1}. We first transform into the interaction picture with respect to $\hat{H}_{p}$ using: 
\begin{eqnarray}
\hat{U}_{p} &=& \exp\Big(-\frac{i}{\hbar}\int^{t}_{0}dt^{\prime}\hat{H}_{p}(t^{\prime}) -\frac{1}{2\hbar^{2}}\int^{t}_{0}\int^{t^{\prime}}_{0}dt^{\prime}dt^{\prime\prime}[\hat{H}_{p}(t^{\prime}),\hat{H}_{p}(t^{\prime\prime})]\Big) \nonumber \\
&=& \exp\Big(\hat{A}[\alpha(t)a^{\dagger}-\alpha^{*}(t)\hat{a}] + i\hat{A}^{2}f(t)\Big),
\end{eqnarray}
where $\alpha(t) \equiv (1-e^{i\Delta t})/\Delta$ and $g(t)\equiv (t-\sin(\Delta t)/\Delta)/\Delta$; the latter commutes with $\hat{H}_{e}$, having no effect on the transformation. The interaction picture Hamiltonian is then:
\begin{eqnarray}\label{eq:interaction_shift}
\hat{H}_{I} &=& \hbar g_{d}\Big(\hat{a}^{\dagger}\hat{a} + \hat{A}[\alpha(t)\hat{a}^{\dagger}+\alpha^{*}(t)\hat{a}] + \hat{A}^{2}|\alpha(t)|^{2}\Big).
\end{eqnarray}
We begin the next section with Eq.~(\ref{eq:interaction_shift}) because $\hat{H}_{e}$ for motional dephasing deviates only through a $ g_{d}\rightarrow  g_{d}\cos(\omega t)$ substitution. We use time-dependent perturbation theory to evaluate the time propagator $\hat{U}_{I}$ for $\hat{H}_{I}$ up to $2^{\text{nd}}$-order:
\begin{eqnarray}
\hat{U}_{I}\simeq \hat{I} - \frac{i}{\hbar}\int^{t_{g}}_{0}\hat{H}_{I}(t^{\prime}) - \frac{1}{\hbar^{2}}\int^{t_{g}}_{0}\int^{t^{\prime}}_{0}dt^{\prime}dt^{\prime\prime}\hat{H}_{I}(t^{\prime})\hat{H}_{I}(t^{\prime\prime}).
\end{eqnarray}
Plugging this into Eq.~(7) in the main text, dropping all terms higher-order than $\mathcal{O}(g_{d}^{2}$), and assuming system is initialized to a state with $n$ phonons, gives:
\begin{eqnarray}
\mathcal{I} &\simeq & \frac{ g_{d}^{2}\Omega_{p}^{2}t_{g}^{2}}{\Delta^{2}}\lambda^{2}_{\hat{\sigma}_{\alpha}}\Big(2n+1 + \frac{16\Omega_{e}^{2}}{\Delta^{2}} \Big) \nonumber \\
&=& \frac{ g_{d}^{2}\theta^{2}}{16\Omega_{e}^{2}}\lambda_{\hat{\sigma}_{\alpha}}^{2}\Big( 2n+1 + \frac{2\theta\Omega_{e}}{\pi \Omega_{p}K}\Big),
\end{eqnarray}
where we have substituted $\Delta \equiv \sqrt{8\pi K\Omega_{p}\Omega_{e}/\theta}$ in the second line.

\subsection*{Motional dephasing}

We take motional dephasing to be the broadband limit of $\hat{H}_{e}$ for static motional frequency shifts. To evaluate the contribution to $\mathcal{I}$, we let $\hat{H}_{e}=\hbar g_{d}\hat{a}^{\dagger}\hat{a}\cos(\omega t)$, calculate the infidelity for that frequency $\mathcal{I}_{\omega}$, then average over the normalized spectral power density $S_{\omega}$. We may begin our evaluation with Eq.~(\ref{eq:interaction_shift}), substituting $ g_{d}\rightarrow g_{d}\cos(\omega t)$, which gives:
\begin{eqnarray}\label{eq:interaction_deph}
\hat{H}_{I,\omega} &=& \hbar g_{d}\cos(\omega t)\Big(\hat{a}^{\dagger}\hat{a} + \hat{A}[\alpha(t)\hat{a}^{\dagger}+\alpha^{*}(t)\hat{a}] + \hat{A}^{2}|\alpha(t)|^{2}\Big).
\end{eqnarray}
Again, we evaluate the interaction picture time propagator using $2^{\text{nd}}$-order time-dependent perturbation theory:
\begin{eqnarray}
\hat{U}_{I}\simeq \hat{I} - \frac{i}{\hbar}\int^{t_{g}}_{0}dt^{\prime}\hat{H}_{I,\omega}(t^{\prime}) - \frac{1}{\hbar^{2}}\int^{t_{g}}_{0}\int^{t^{\prime}}_{0}dt^{\prime}dt^{\prime\prime}\hat{H}_{I,\omega}(t^{\prime})\hat{H}_{I,\omega}(t^{\prime\prime}).
\end{eqnarray}
Plugging this into Eq.~(7) of the main text, and dropping all terms higher-order than $\mathcal{O}(g_{d}^{2})$ gives:
\begin{eqnarray}
\mathcal{I}_{\omega}&\simeq & \frac{2}{\hbar^{2}}\sum_{n^{\prime}}\int^{t_{g}}_{0}\int^{t^{\prime}}_{0}dt^{\prime}dt^{\prime\prime}\bra{\psi(0)}\bra{n^{\prime}}\hat{H}_{I,\omega}(t^{\prime})\hat{H}_{I,\omega}(t^{\prime\prime})\ket{\psi(0)}\ket{n} \nonumber \\
&& - \frac{1}{\hbar^{2}}\sum_{n^{\prime}}\int^{t_{g}}_{0}\int^{t_{g}}_{0}dt^{\prime}dt^{\prime\prime}\bra{\psi(0)}\bra{n^{\prime}}\hat{H}_{I,\omega}(t^{\prime})\ket{\psi(0)}\ket{n}\bra{\psi(0)}\bra{n}\hat{H}_{I,\omega}(t^{\prime\prime})\ket{\psi(0)}\ket{n^{\prime}}
.
\end{eqnarray}
We calculate $\mathcal{I}$ by integrating over $S_{\omega}$, assuming it is broad enough to warrant the Born-Markov approximation:
\begin{eqnarray}\label{eq:infid_ham_only_mot}
\mathcal{I} &=& \int^{\infty}_{0}d\omega S_{\omega}\mathcal{I}_{\omega} \nonumber \\
&\simeq & \frac{2}{\hbar^{2}}\sum_{n^{\prime}}\int^{\infty}_{0}\int^{t_{g}}_{0}\int^{t^{\prime}}_{0}d\omega dt^{\prime}dt^{\prime\prime}S_{\omega}\bra{\psi(0)}\bra{n^{\prime}}\hat{H}_{I,\omega}(t^{\prime})\hat{H}_{I,\omega}(t^{\prime\prime})\ket{\psi(0)}\ket{n} \nonumber \\
&& - \frac{1}{\hbar^{2}}\sum_{n^{\prime}}\int^{\infty}_{0}\int^{t_{g}}_{0}\int^{t_{g}}_{0}d\omega dt^{\prime}dt^{\prime\prime}S_{\omega}\bra{\psi(0)}\bra{n^{\prime}}\hat{H}_{I,\omega}(t^{\prime})\ket{\psi(0)}\ket{n}\bra{\psi(0)}\bra{n}\hat{H}_{I,\omega}(t^{\prime\prime})\ket{\psi(0)}\ket{n^{\prime}}.
\end{eqnarray}
Similar to Ref.~\cite{sutherland_2022_1}, upon plugging Eq.~(\ref{eq:interaction_deph}) into Eq.~(\ref{eq:infid_ham_only_mot}), we are left with a sum of triple integrals that are proportional to:
\begin{eqnarray}\label{eq:integrals_mot_deph}
\zeta &=& \!\int^{\infty}_{0}\!\!\int^{t_{g}}_{0}\!\int^{t_{s}}_{0}\! d\omega dt^{\prime}dt^{\prime\prime}\frac{S_{\omega}}{2}\Big\{\!\cos(\omega[t^{\prime\prime}\!+\!t^{\prime}]) \!+\!\cos(\omega[t^{\prime\prime}\!-\!t^{\prime}])\Big\}, \nonumber \\
\end{eqnarray}
where $t_{s}\in\{t_{g},t^{\prime}\}$. In order to evaluate Eq.~(\ref{eq:integrals_mot_deph}), we perform the following manipulations:
\begin{eqnarray}
\zeta &=& \!\int^{\infty}_{-\infty}\!\int^{t_{g}}_{0}\!\int^{t_{s}}_{0}d\omega dt^{\prime}dt^{\prime\prime}\frac{S_{\omega}}{4}\Big\{\!\cos(\omega[t^{\prime\prime}\! +\! t^{\prime}]) \!+\!\cos(\omega[t^{\prime\prime}\!-\! t^{\prime}])\!\Big\} \nonumber \\
&\simeq& \frac{S_{0}}{4}\!\int^{\infty}_{-\infty}\!\int^{t_{g}}_{0}\!\int^{t_{s}}_{0}\! d\omega  dt^{\prime} dt^{\prime\prime}\!\Big\{\!\cos(\omega[t^{\prime\prime}\!+\! t^{\prime}]) \!+\!\cos(\omega[t^{\prime\prime}\!-\! t^{\prime}])\Big\} \nonumber \\
&=& \frac{\pi S_{0}}{2}\int^{t_{g}}_{0}\int^{t_{s}}_{0} dt^{\prime}dt^{\prime\prime}\Big\{\delta(t^{\prime\prime}+t^{\prime}) + \delta(t^{\prime\prime}-t^{\prime})\Big\},
\end{eqnarray}
where we have assumed that we can approximate $S_{\omega}$ as $S_{0}$, pulling it ouside the integral. We can now evaluate the required integrals, and, after some algebra, obtain a final equation:
\begin{eqnarray}
\mathcal{I} &\simeq & \frac{2\eta\Omega_{p}^{2} t_{g}}{\Delta^{2}}\Big(2n+1 + \frac{12\Omega_{e}^{2}}{\Delta^{2}}\Big)\lambda_{\hat{\sigma}_{\alpha}}^{2} \nonumber \\
&= & \eta\sqrt{\frac{\Omega_{p}\theta^{3}}{32\pi K \Omega_{e}^{3}}}\Big(2n+1 + \frac{3\theta\Omega_{e}}{2\pi K\Omega_{p}} \Big)\lambda_{\hat{\sigma}_{\alpha}}^{2},
\end{eqnarray}
where we have defined $\eta \equiv \pi g_{d}^{2}S_{0}/2$ as the motional dephasing rate, and set $\Delta \equiv \sqrt{8\pi K\Omega_{p}\Omega_{e}/\theta}$ as well as $t_{g} = 2\pi K/\Delta$ in the second line.

\subsection*{Heating}

Finally, we discuss how heating affects gate fidelity. As in Ref.~\cite{sutherland_2022_1}, we represent $\hat{H}_{e}$ as an E-field with frequency $\omega$:
\begin{eqnarray}
\hat{H}_{t} &=& \hat{H}_{p} + \hat{H}_{e} \nonumber \\
&=& \hbar\hat{A}\Big(\hat{a}^{\dagger}e^{i \Delta t} + \hat{a}e^{-i\Delta t}\Big) + 2\hbar  g_{h}\cos(\omega t)\Big(\hat{a}^{\dagger}e^{i\omega_{r}t} + e^{-i\omega_{r}t} \Big) \nonumber \\
&\simeq & \hbar\hat{A}\Big(\hat{a}^{\dagger}e^{i \Delta t} + \hat{a}e^{-i\Delta t}\Big) + \hbar  g_{h}\Big(\hat{a}^{\dagger}e^{i(\omega_{r}-\omega)t} + e^{-i(\omega_{r}-\omega)t} \Big),
\end{eqnarray}
where we have made made the rotating wave approximation in the second line. We now transform into the interaction picture with respect to $\hat{H}_{e}$ using the transformation:
\begin{eqnarray}
\hat{U}_{e} = \exp\Big(\beta(t)\hat{a}^{\dagger} - \hat{\beta}^{*}(t)\hat{a} \Big), 
\end{eqnarray}
where $\beta(t) \equiv \frac{ g_{h}}{\omega_{r}-\omega}(1 - e^{i[\omega_{r}-\omega]t})$. This gives:
\begin{eqnarray}
\hat{H}_{I} &=& \hbar\hat{A}\Big(\hat{a}^{\dagger}e^{i \Delta t} + \hat{a}e^{-i\Delta t}\Big) + \hbar\hat{A}\Big(\beta(t)^{*}e^{i\Delta t} + \beta(t) e^{-i\Delta t}\Big) \nonumber \\
&\equiv & \hat{H}_{I,p} + \hat{H}_{I,e}
\end{eqnarray}
showing the effect of the extraneous E-field is an added $\propto \hat{A}$ shift to the Hamiltonian in this frame. Importantly, $\hat{H}_{I,p}=\hat{H}_{p}$ and $[\hat{H}_{I,p},\hat{H}_{I,e}]=0$. This allows us to write the time propagator as $\hat{U}_{p}\hat{U}_{I,e}$, where:
\begin{eqnarray}
\hat{U}_{I,e} = \exp\Big(-\frac{i}{\hbar}\int^{t_{g}}_{0}dt^{\prime}\hat{H}_{I,e}(t^{\prime}) \Big). 
\end{eqnarray}
We Taylor expand $\hat{U}_{I,e}$ up to $2^{\text{nd}}$-order, and plug it into Eq.~(7) of the main text, which gives: 
\begin{eqnarray}
\mathcal{I}_{\omega}&\simeq & 1-\sum_{n^{\prime}}\Big|\bra{\psi(0)}\bra{n^{\prime}}\hat{I} - i\hat{A}\int^{t_{g}}_{0}dt^{\prime}\Big(\beta^{*}(t^{\prime})e^{i\Delta t^{\prime}} + \beta(t^{\prime})e^{-i\Delta t^{\prime}}\Big) \nonumber \\
&& - \frac{\hat{A}^{2}}{2}\int^{t_{g}}_{0}\int^{t^{\prime}}_{0}dt^{\prime}dt^{\prime\prime}\Big(\beta^{*}(t^{\prime})e^{i\Delta t^{\prime}} + \beta(t^{\prime})e^{-i\Delta t^{\prime}}\Big)\Big(\beta^{*}(t^{\prime\prime})e^{i\Delta t^{\prime\prime}} + \beta(t^{\prime\prime})e^{-i\Delta t^{\prime\prime}}\Big)\ket{\psi(0)}\ket{n}\Big|^{2} \nonumber \\
&=& \lambda^{2}_{\hat{A}}\int^{t_{g}}_{0}\int^{t_{g}}_{0}dt^{\prime}dt^{\prime\prime}\Big(\beta^{*}(t^{\prime})e^{i\Delta t^{\prime}} + \beta(t^{\prime})e^{-i\Delta t^{\prime}}\Big)\Big(\beta^{*}(t^{\prime\prime})e^{i\Delta t^{\prime\prime}} + \beta(t^{\prime\prime})e^{-i\Delta t^{\prime\prime}}\Big).
\end{eqnarray}
We can plug this into Eq.~(8) from the main text, and integrate over the normalized spectral density $S_{\omega}$:
\begin{eqnarray}
\mathcal{I} &=& \lambda^{2}_{\hat{A}}\int^{\infty}_{0}\int^{t_{g}}_{0}\int^{t_{g}}_{0}d\omega dt^{\prime}dt^{\prime\prime} S_{\omega}\frac{4 g_{h}^{2}}{(\omega_{r}-\omega)^{2}}\Big(\cos(\Delta t^{\prime})  -\cos\Big[(\omega_{r}-\omega-\Delta)t^{\prime}\Big] \Big)\Big(\cos(\Delta t^{\prime\prime})  -\cos\Big[(\omega_{r}-\omega-\Delta)t^{\prime\prime}\Big]  \Big) \nonumber \\
&=&\lambda^{2}_{\hat{A}}\int^{\infty}_{0}\int^{t_{g}}_{0}\int^{t_{g}}_{0}d\omega dt^{\prime}dt^{\prime\prime} S_{\omega}\frac{4 g_{h}^{2}}{(\omega_{r}-\omega)^{2}}\cos([\omega_{r}-\omega-\Delta]t^{\prime})\cos([\omega_{r}-\omega-\Delta]t^{\prime\prime}) \nonumber \\
&\simeq&\lambda^{2}_{\hat{A}}\int^{\infty}_{0}\int^{t_{g}}_{0}\int^{\infty}_{-\infty}d\omega dt^{\prime}dt^{\prime\prime} S_{\omega}\frac{2  g_{h}^{2}}{(\omega_{r}-\omega)^{2}}\cos([\omega_{r}-\omega-\Delta]t^{\prime}) \cos([\omega_{r}-\omega-\Delta]t^{\prime\prime}) \nonumber \\
&=& \lambda^{2}_{\hat{A}}\int^{\infty}_{0}\int^{t_{g}}_{0}d\omega dt^{\prime} S_{\omega}\frac{4\pi  g_{h}^{2}}{(\omega_{r}-\omega)^{2}}\cos([\omega_{r}-\omega-\Delta]t^{\prime})\delta(\omega_{r}-\omega-\Delta) \nonumber \\
&=& \frac{4\dot{\bar{n}}t_{g}}{\Delta^{2}}\lambda_{\hat{A}}^{2},
\end{eqnarray}
where we have substituted $\dot{\bar{n}}\equiv \pi  g_{h}^{2}S_{\omega_{r}}$ in the last line. Finally, we can substitute $\Delta \equiv \sqrt{8\pi K\Omega_{p}\Omega_{e}/\theta}$ and $\lambda^{2}_{\hat{A}}=\Omega_{p}^{2}\lambda^{2}_{\hat{\sigma}_{\alpha}}$, giving:
\begin{eqnarray}
\mathcal{I} \simeq \dot{\bar{n}}\lambda^{2}_{\hat{\sigma}_{\alpha}}\sqrt{\frac{\Omega_{p}\theta^{3}}{8\pi K\Omega_{e}^{3}}}. 
\end{eqnarray}

\clearpage

\end{document}